\documentclass[11pt,a4paper]{article}
\usepackage{amsmath}
\usepackage{amsfonts}

\input{tcilatex}
\begin{document}

\title{SWKB and proper quantization conditions for translationally shape
invariant potentials }
\author{Kamal Mahdi$^{1}$,Y. Kasri$^{1,2}$, Y. Grandati$^{2}$ and A. B\'{e}%
rard$^{2}$ \and $^{1}$ \textit{Laboratoire de Physique Th\'{e}orique, Facult%
\'{e} des Sciences Exactes, } \and \textit{Universit\'{e} de Bejaia, 06000
Bejaia, Alg\'{e}rie.} \and $^{2}$\textit{\ Equipe BioPhysStat, Laboratoire
LCP-A2MC, ICPMB, IF CNRS \ N}$^{\text{0}}$\textit{2843,} \and \textit{\
Universit\'{e} de Lorraine, 1 bvd D.F. Arago, 57078 Metz, France.}}
\date{}
\maketitle

\begin{abstract}
Using a recently proposed classification for the primary translationally
shape invariant potentials, we show that the exact quantization rule
formulated by Ma and Xu is equivalent to the supersymmetric JWKB
quantization condition. The energy levels for the two considered categories
of shaped invariant potentials are also derived.
\end{abstract}

\baselineskip=1.0\baselineskip

Recently, a new quantization rule in quantum mechanics which leads to the
exact spectrum for some exactly solvable systems was proposed by Ma and Xu 
\cite{1}-\cite{2}. This rule brings a correction term to the well known JWKB
quantization condition. This quantum correction is given in the form of an
integral, which contains the solution of Riccati's equation associated to
the Schr\"{o}dinger's equation for the ground state. Qiang and Dong \cite{3}
have simplified the formulation of Ma-Xu rule, calling it the proper
quantization rule. Since its original formulation by Ma and Xu in 2005,
several papers have been published on its possible applications to different
potentials. This rule has been applied successfully to numerous solvable
potentials in one and three dimensions, as well as separable non-central
potentials \cite{4}. In fact it has been proven that the Ma-Xu formula is
necessary exact for every primary translationally shape invariant potential
(PTSIP) \cite{5}. As recently discovered , there exist also infinite towers
of secondary TSIP for which the Ma-Xu quantization rule has not been proven
to be exact \cite{52,53a,53b}.

Another quantization rule based on supersymmetric quantum mechanics, the
so-called SWKB rule \cite{6}-\cite{8}, permits the calculation of energy
levels. It is also exact for every PTSIP but not for secondary TSIP \cite{81}%
.

Until now, the question about a direct link between the two quantization
rules remains still unclear. The main purpose of this paper is to examine
this link. Adopting the classification into two categories of PTSIP
introduced in \cite{9}, we show that, for these potentials, the proper
quantization rule is in fact equivalent to the SWKB condition. Starting from
the SWKB condition, by means of some transformations and using appropriate
changes of variables, we derive the proper quantization formula and then
establish a direct and explicit connection between the two quantization
rules in the class of potentials for which they appear to be exact. The
quantization formulas allow then to obtain global formulas for the energy
spectrum of the two categories of PTSIP and we apply it to retrieve the
energy levels of the generic PTSIP.

The paper is organised as follows: the classification into two categories of
potentials is introduced in the first section, then a brief introduction of
the two quantization formulas is presented in the second section. In the
next paragraph, the connection between these two rules is established. As an
application, we calculate the energy levels for the two considered
categories of potentials in the last section.

\section{Potentials of first and second category}

Based on an analysis of the structure of PTSIP, Grandati and Berard \cite{9}
have shown that these potentials can be classified into two categories,
simplifying the preceding classifications proposed initially by Gendenshtein 
\cite{Gendenshtein} and later by Barclay \cite{barclay2}. In what follows,
we use this classification. The basic idea is that every PTSIP can be
reduced into a second degree polynomial or a second degree Laurent
polynomial via a uniquely defined change of variable.

\subsection{First category (generic)}

We say that a one-dimensional potential $V(x)$ is a generic potential of the
first category \cite{9} if, with a suitable choice of a function $y(x)$, the
two following conditions on the potential and $y$ are fulfilled

\begin{equation}
\left\{ 
\begin{array}{c}
V(x)\rightarrow V(y)=\lambda _{2}y^{2}+\lambda _{1}y+\lambda _{0}\text{,\ }%
\lambda _{2}>0 \\ 
y^{\prime }(x)=\alpha (1\pm y^{2})>0.\text{ \ \ \ \ \ \ \ \ \ \ \ \ \ \ \ \
\ \ \ \ \ \ \ \ \ }%
\end{array}%
\right.  \label{7}
\end{equation}

In this category we find the following potentials : Rosen-Morse I,
Rosen-Morse II and Eckart.\textbf{\ }It also contains the three exceptional
cases corresponding to the harmonic, Morse and effective radial
Kepler--Coulomb potentials.

We now seek the superpotential associated to this category. For a particle
subjected to a potential $V(x)$ with a bound state spectrum $\left(
E_{n},\psi _{n}\right) $, the superpotential is given in terms of the ground
state wave function $\psi _{0}$ by the relation (in units $\hbar =2m=1$) 
\begin{equation}
W(x)=-\frac{d}{dx}\left( \ln \psi _{0}\right) ,  \label{1}
\end{equation}%
$W(x)$ being a solution of the nonlinear Riccati equation

\begin{equation}
W^{2}-W^{\prime }(x)=V(x)-E_{0},  \label{3}
\end{equation}%
obtained by inserting $\left( \ref{1}\right) $ into the Schr\"{o}dinger
equation $\psi _{n}^{\prime \prime }(x)+\left( E_{n}-V(x)\right) \psi
_{n}(x)=0$ for $n=0$. Looking for a polynomial solution $W\left( y\right) $
of $\left( \ref{3}\right) $ expressed in the $y$ variable, we obtain \cite{9}%
: $W\left( y\right) =ay+b$ $\left( a>0\right) $, where the coefficients $a$
and $b$ and the ground state energy $E_{0}$ are given by

\begin{eqnarray}
a &=&\pm \frac{\alpha }{2}+\sqrt{\left( \alpha /2\right) ^{2}+\lambda _{2}},%
\text{ }b=\frac{\lambda _{1}}{2a},  \label{9} \\
E_{0} &=&\lambda _{0}+\alpha a-b^{2}.  \label{10}
\end{eqnarray}

From the first equation, we get a relation which will be useful later%
\begin{equation}
a(a\mp \alpha )=\lambda _{2}.  \label{11}
\end{equation}

\subsection{Second category (generic)}

We say that a one-dimensional potential is a generic potential of second
category \cite{9} if the two following conditions are satisfied

\begin{equation}
\left\{ 
\begin{array}{c}
V(x)\rightarrow V(y)=\lambda _{2}y^{2}+\frac{\mu _{2}}{y^{2}}+\lambda _{0},%
\text{ \ \ }\lambda _{2},\mu _{2}>0, \\ 
y^{\prime }(x)=\alpha (1\pm y^{2})>0.\text{\ \ \ \ \ \ \ \ \ \ \ \ \ \ \ \ \
\ \ \ \ \ \ \ \ \ \ \ \ \ \ }%
\end{array}%
\right.  \label{15.1}
\end{equation}

This category contains the isotonic, P\"{o}schl-Teller, P\"{o}schl-Teller I,
P\"{o}schl-Teller II and Scarf \ potentials. Looking for an associated
superpotential $W\left( y\right) $ solution of \ the Riccati equation in the 
$y$ variable, which is a Laurent polynomial, we obtain \cite{9}: $W\left(
y\right) =ay-\frac{b}{y}$ $\left( a>0,b>0\right) $ where the unknown
coefficients $a$ and $b$ and the ground state energy $E_{0}$ are then
obtained as%
\begin{eqnarray}
a &=&\pm \frac{\alpha }{2}+\sqrt{\left( \alpha /2\right) ^{2}+\lambda _{2}}%
,b=\frac{\alpha }{2}+\sqrt{\left( \alpha /2\right) ^{2}+\mu _{2}},
\label{15.2} \\
E_{0} &=&\lambda _{0}+2ab+\alpha \left( a\pm b\right) .  \label{15.3}
\end{eqnarray}

From $\left( \ref{15.2}\right) $, we get the two useful relations

\begin{equation}
a(a\mp \alpha )=\lambda _{2},\text{ }b(b-\alpha )=\mu _{2}.  \label{15.4}
\end{equation}

\section{The quantization rules}

\subsection{The SWKB condition}

The SWKB quantization condition \cite{8} is the supersymmetric version of
the well known JWKB condition and allows to estimate the energy levels of $%
V(x)$ via the identity%
\begin{equation}
I^{SWKB}=\int_{x_{L}}^{x_{R}}\sqrt{E_{n}^{(-)}-W^{2}(x)}dx=n\pi ,  \label{4}
\end{equation}%
where $x_{SL}$\ and $x_{SR}$ are the SWKB turning points determined by $%
E_{n}^{(-)}=W^{2}\left( x\right) $ and $E_{n}^{(-)}$ is related to the
particle energy by\ $E_{n}^{(-)}=E_{n}-E_{0}$. Interestingly, the SWKB
reproduces the exact energy spectrum for the complete set of PTSIP \cite{8},
which contains all the classical exactly solvable quantum potentials in
closed analytical form.

In the following, we note $y_{SR}$ and $y_{SR}$, the $y$ turning points
corresponding to the SWKB condition. After straightforward calculations, it
results that they are determined, for the first category, by%
\begin{equation}
y_{SR}+y_{SL}=-\frac{E_{n}^{(-)}-b^{2}}{a^{2}}\text{, \ \ \ }y_{SR}.y_{SL}=-%
\frac{2b}{a}\text{,}  \label{14}
\end{equation}%
consequently, the integral in $\left( \ref{4}\right) $ can be rewritten as%
\begin{equation}
I^{SWKB}=\frac{a}{\alpha }\dint_{y_{SL}}^{y_{SR}}\sqrt{\left(
y_{SR}-y\right) \left( y-y_{SL}\right) }\frac{dy}{1\pm y^{2}}.  \label{15}
\end{equation}

For the second category and with the help of a second change of variable $%
z=y^{2}$, the $z$ turning points are

\begin{equation}
z_{SR}+z_{SL}=\frac{E_{n}^{(-)}+2ab}{a^{2}}\text{, }z_{SR}.z_{SL}=\frac{b^{2}%
}{a^{2}},  \label{15.5}
\end{equation}%
and for the integral in $\left( \ref{4}\right) $, we obtain%
\begin{equation}
I^{SWKB}=\frac{a}{2\alpha }\dint_{z_{SL}}^{z_{SR}}\sqrt{\left(
z_{SR}-z\right) \left( z-z_{SL}\right) }\frac{dz}{z\left( 1\pm z\right) },
\label{15.6}
\end{equation}

\subsection{The Ma-Xu or proper quantization formula}

A few years ago, a semi-classical quantization rule was proposed by Ma and
Xu \cite{1,2} 
\begin{equation}
\int_{x_{L}}^{x_{R}}\sqrt{E_{n}-V(x)}dx=n\pi +\gamma .  \label{5}
\end{equation}

This rule generalizes the JWKB condition by the addition of a term $\gamma $
which takes into account the higher order quantum corrections and which
appears to be necessarily energy independent for all the PTSIP \cite{5}. It
then gives a new way to access to the exact energy spectrum for this exactly
solvable potential. In practice, it is first necessary to determine the
energy $E_{0}$ of the ground state by solving the Riccati equation $\left( %
\ref{3}\right) $, then proceed to the evaluation of the integral $%
I^{PQ}\left( E_{n}\right) $ and replace $n=0$ to obtain $I^{PQ}\left(
E_{0}\right) $.

Later Qiang and Dong have proposed a slightly improved version of the Ma-Xu
formula, they called the proper quantization rule, which is given under the
following form 
\begin{equation}
I^{PQ}\left( E_{n}\right) =I^{PQ}\left( E_{0}\right) +n\pi ,  \label{6}
\end{equation}%
where $I^{PQ}\left( E_{n}\right) $ is the integral in the left hand side of
equation $\left( \ref{5}\right) $.

Let us now determine the$\ y$ turning points corresponding to proper
quantization formula. For the first category, one obtains%
\begin{equation}
y_{PR}+y_{PL}=-\frac{E_{n}-\lambda _{0}}{\lambda _{2}}\text{, \ \ \ \ \ }%
y_{PR}.y_{PL}=-\frac{\lambda _{1}}{\lambda _{2}}\text{,}  \label{6.1}
\end{equation}%
and the integral $\left( \ref{5}\right) $ can be rewritten as%
\begin{equation}
I^{PQ}\left( E_{n}\right) =\frac{\sqrt{\lambda _{2}}}{\alpha }%
\dint_{y_{PL}}^{y_{PR}}\sqrt{\left( y_{PR}-y\right) \left( y-y_{PL}\right) }%
\frac{dy}{1\pm y^{2}}.  \label{6.2}
\end{equation}

For the second category, again with the help of a second change of variable $%
z=y^{2}$, the $z$ turning points are given by

\begin{equation}
z_{PR}+z_{PL}=\frac{E_{n}-\lambda _{0}}{\lambda _{2}}\text{, }z_{PR}.z_{PL}=%
\frac{\mu _{2}}{\lambda _{2}},  \label{15.7}
\end{equation}%
and the integral in $\left( \ref{5}\right) $ becomes%
\begin{equation}
I^{PQ}\left( E_{n}\right) =\frac{\sqrt{\lambda _{2}}}{2\alpha }%
\dint_{z_{PL}}^{z_{PR}}\sqrt{\left( z_{PR}-z\right) \left( z-z_{PL}\right) }%
\frac{dz}{z\left( 1\pm z\right) }.  \label{15.8}
\end{equation}

\section{From SWKB to proper quantization rule}

\subsection{First category}

We shall now investigate the relationship between $I^{SWKB}\left(
E_{n}\right) $ and $I^{PQ}\left( E_{n}\right) $ for the first category of
potentials. The presence of the symbol $\pm $ \ in $\left( \ref{15}\right) $
means that we have in fact to evaluate two distinct integrals.

Starting from $\left( \ref{15}\right) $ and making use of the integral
formulas $\left( \ref{13a}\right) $ and $\left( \ref{13b}\right) $ combined
with expressions of the turning points given by $\left( \ref{14}\right) $,
we obtain

\begin{equation}
I^{SWKB}=\left\{ 
\begin{array}{c}
\frac{\pi }{\sqrt{2}\alpha }\left( \sqrt{\left( E_{n}^{-}+a^{2}-b^{2}\right)
^{2}+4a^{2}b^{2}}+E_{n}^{-}+a^{2}-b^{2}\right) ^{\frac{1}{2}}-\frac{\pi a}{%
\alpha }\text{, \ if }y^{\prime }=\alpha \left( 1+y^{2}\right) , \\ 
\\ 
\frac{\pi }{2\alpha }\left[ 2a-\sqrt{\left( a+b\right) ^{2}-E_{n}^{-}}-\sqrt{%
\left( a-b\right) ^{2}-E_{n}^{-}}\right] \text{ \ \ \ \ \ \ \ \ \ \ \ \ \ \
\ ,\ if }y^{\prime }=\alpha \left( 1-y^{2}\right) .%
\end{array}%
\right.  \label{17}
\end{equation}

In order to involve $I^{PQ}\left( E_{n}\right) $, we need to express the
previous integrals in terms of $\lambda _{1}$, $\lambda _{2}$\ and $E_{n}$.

In the first case, i.e. when $y^{\prime }\left( x\right) =\alpha \left(
1+y^{2}\right) $, using equations $\left( \ref{9}\right) $ to $\left( \ref%
{11}\right) $ and $E_{n}^{(-)}=E_{n}-E_{0}$, it is straightforward to show
that 
\begin{equation}
\left( E_{n}^{-}+a^{2}-b^{2}\right) +^{2}4a^{2}b^{2}=\left( E_{n}-\lambda
_{0}+\lambda _{2}\right) ^{2}+\lambda _{1}^{2}.  \label{18}
\end{equation}

In the second case, i.e. when $y^{\prime }\left( x\right) =\alpha \left(
1-y^{2}\right) $, we obtain in the same way 
\begin{equation}
\left( a\pm b\right) ^{2}-E_{n}^{-}=\lambda _{2}+\lambda _{0}-E_{n}\mp
\lambda _{1}.  \label{19}
\end{equation}

Inserting the two previous equations into $\left( \ref{17}\right) $, and
after a little algebra, one finds

\begin{equation}
I^{SWKB}=\left\{ 
\begin{array}{c}
\frac{\pi \sqrt{\lambda _{2}}}{\alpha }\left[ \frac{1}{\sqrt{2}}\left( \sqrt{%
\left( \frac{E_{n}-\lambda _{0}}{\lambda _{2}}+1\right) ^{2}+\left( \frac{%
\lambda _{1}}{\lambda _{2}}\right) ^{2}}+\frac{E_{n}-\lambda _{0}}{\lambda
_{2}}+1\right) ^{\frac{1}{2}}-1\right] -\frac{\pi }{\alpha }\left( a-\sqrt{%
\lambda _{2}}\right) \\ 
\\ 
\frac{\pi \sqrt{\lambda _{2}}}{2\alpha }\left[ 2-\sqrt{1+\frac{\lambda _{1}}{%
\lambda _{2}}-\frac{E_{n}-\lambda _{0}}{\lambda _{2}}}-\sqrt{1-\frac{\lambda
_{1}}{\lambda _{2}}-\frac{E_{n}-\lambda _{0}}{\lambda _{2}}}\right] +\frac{%
\pi }{\alpha }\left( a-\sqrt{\lambda _{2}}\right) \text{.\ \ \ \ \ \ \ \ \ \
\ \ \ }%
\end{array}%
\right.  \label{20}
\end{equation}

Using relations $\left( \ref{6.2}\right) $ and formulas $\left( \ref{13a}%
\right) $ and $\left( \ref{13b}\right) $, one can rewrite $\left( \ref{20}%
\right) $ as

\begin{equation}
I^{SWKB}=\frac{\sqrt{\lambda _{2}}}{\alpha }\dint_{y_{PL}}^{y_{PR}}\sqrt{%
\left( y_{PR}-y\right) \left( y-y_{PL}\right) }\frac{dy}{1\pm y^{2}}-\left(
\pm \frac{\pi }{\alpha }\left( a-\sqrt{\lambda _{2}}\right) \right) .
\label{21}
\end{equation}

Then, taking into account the fact that (see Appendix 1 for details)%
\begin{equation}
I^{PQ}\left( E_{0}\right) =\pm \frac{\pi }{\alpha }\left( a-\sqrt{\lambda
_{2}}\right)  \label{22}
\end{equation}%
and using $\left( \ref{6.2}\right) $, equation $\left( \ref{21}\right) $
becomes%
\begin{equation}
I^{SWKB}=I^{PQ}\left( E_{n}\right) -I^{PQ}\left( E_{0}\right) ,  \label{23}
\end{equation}%
which means that the SWKB condition is identical to the proper quantization
rule for the first category of PTSIP.

\subsection{Second category}

Starting from $\left( \ref{15.6}\right) $ and with the help of the integral
formula $\left( \ref{15c}\right) $ combined to expressions of the turning
points given by $\left( \ref{15.5}\right) $, $I^{SWKB}$ can be written as%
\begin{equation}
I^{SWKB}=\frac{\pi }{2\alpha }\left( \mp a-b\pm \sqrt{\left( b\pm a\right)
^{2}\pm E_{n}^{-}}\right) .  \label{170}
\end{equation}

In order to make appear explicitly $I^{PQ}\left( E_{n}\right) $ in this
expression, we can use the following identity (see $\left( \ref{15.2}\right) 
$ to $\left( \ref{15.4}\right) $)

\begin{equation}
\left( b\pm a\right) ^{2}\pm E_{n}^{-}=\pm E_{n}\mp \lambda _{0}+\lambda
_{2}+\mu _{2}.  \label{180}
\end{equation}

Inserting $\left( \ref{180}\right) $ into $\left( \ref{170}\right) $ leads to

\begin{equation}
I^{SWKB}=\frac{\sqrt{\lambda _{2}}}{2\alpha }\pi \left[ \left( \mp 1-\sqrt{%
\frac{\mu _{2}}{\lambda _{2}}}\pm \sqrt{1+\frac{\mu _{2}}{\lambda _{2}}\pm 
\frac{E_{n}-\lambda _{0}}{\lambda _{2}}}\right) -\left( \frac{b\pm a}{\sqrt{%
\lambda _{2}}}\mp 1-\sqrt{\frac{\mu _{2}}{\lambda _{2}}}\right) \right] .
\label{200}
\end{equation}

Refering to $\left( \ref{15.7}\right) $ and to the integral formulas $\left( %
\ref{15c}\right) $, one finds

\begin{equation}
I^{SWKB}=\frac{\sqrt{\lambda _{2}}}{2\alpha }\dint_{z_{PL}}^{z_{PR}}\sqrt{%
\left( z_{PR}-z\right) \left( z-z_{PL}\right) }\frac{dz}{z\left( 1\pm
z\right) }-\frac{\sqrt{\lambda _{2}}}{2\alpha }\left( \frac{b\pm a}{\sqrt{%
\lambda _{2}}}\mp 1-\sqrt{\frac{\mu _{2}}{\lambda _{2}}}\right) .
\label{220}
\end{equation}

But we have (see Appendix 1) 
\begin{equation}
I^{PQ}\left( E_{0}\right) =\frac{\pi }{2\alpha }\left( b\pm a\mp \sqrt{%
\lambda _{2}}-\sqrt{\mu _{2}}\right) ,  \label{230}
\end{equation}%
which leads to the expected result%
\begin{equation}
I^{SWKB}=I^{PQ}\left( E_{n}\right) -I^{PQ}\left( E_{0}\right) ,  \label{240}
\end{equation}%
and proves that the two quantization rules are also equivalent for the
second category PTSIP.

\section{Exceptional potentials}

In order to complete the study of the equivalence between the two
quantization conditions, we still have to examine the four exceptional PTSIP 
\cite{9} which necessitate a specific treatment. In the first category, the
exceptional PTSIP are the harmonic potential, the Morse potential and
Kepler-Coulomb potential. The isotonic potential is the unique exceptional
potential belonging to the second category. In this section, the ground
state energy $E_{0}$\ is taken equal to zero.

\subsection{One-dimensional harmonic oscillator}

The one-dimensional harmonic potential with zero-energy ground state is \cite%
{6}, \cite{11}%
\begin{equation}
V(x)=\frac{\omega ^{2}}{4}x^{2}-\frac{\omega }{2}
\end{equation}%
and the corresponding superpotential is \cite{9} : $W\left( x\right) =\omega
x/2$ . The SWKB turning points are $x_{SR}=-x_{SL}=\left( 2/\omega \right) 
\sqrt{E_{n}^{(-)}}$ and those corresponding to the proper quantization are $%
x_{PR}=-x_{PL}=\left( 2/\omega \right) \sqrt{E_{n}+\omega /2}$. The
evaluation of the $I^{SWKB}$ gives%
\begin{equation}
I^{SWKB}=\frac{\omega }{2}\int_{x_{SL}}^{x_{SR}}\sqrt{(x_{SR}-x)\left(
x-x_{SL}\right) }dx=\frac{\omega }{2}\left( \frac{\pi }{8}\left(
x_{SR}-x_{SL}\right) ^{2}\right) .
\end{equation}

It is straightforward to prove that $\left( x_{SR}-x_{SL}\right) ^{2}=\left(
x_{PR}-x_{PL}\right) ^{2}-\left( x_{PR0}-x_{PL0}\right) ^{2}$ and since 
\begin{equation}
I^{PQ}\left( E_{n}\right) =\frac{\omega }{2}\int_{x_{PL}}^{x_{PR}}\sqrt{%
(x_{PR}-x)\left( x-x_{PL}\right) }dx,
\end{equation}%
we deduce immediately%
\begin{equation}
I^{SWKB}=I^{PQ}\left( E_{n}\right) -I^{PQ}\left( E_{0}\right) .
\end{equation}

\subsection{Morse potential}

The Morse potential with zero-energy ground state is \cite{6}, \cite{11}%
\begin{equation}
V(x)=A^{2}+B^{2}e^{-2\alpha x}-2B(A+\frac{\alpha }{2})e^{-\alpha x},
\end{equation}%
where the parameters $A,B$ and$\ \alpha \ $are all positive. The
introduction of the variable $y=e^{-\alpha x}$ transforms this potential
into $V(y)=\left( A-By\right) ^{2}-\alpha By$ and the corresponding
superpotential is given by \cite{9} : $W\left( y\right) =-By+A$. The turning
points corresponding to the SWKB are determined by 
\begin{equation}
\left\{ 
\begin{array}{c}
y_{SR}+y_{SL}=2A/B, \\ 
y_{SR}.y_{SL}=\left( A^{2}-E_{n}^{(-)}\right) /B^{2}%
\end{array}%
\right.
\end{equation}%
and those for the proper quantization by 
\begin{equation}
\left\{ 
\begin{array}{c}
y_{PR}+y_{PL}=\left( 2A+\alpha \right) /B, \\ 
y_{PR}.y_{PL}=\left( A^{2}-E_{n}\right) /B^{2}.%
\end{array}%
\right.
\end{equation}

We have%
\begin{equation}
I^{SWKB}=\frac{B}{\alpha }\dint_{y_{SL}}^{y_{SR}}\sqrt{\left(
y_{SR}-y\right) \left( y-y_{SL}\right) }\frac{dy}{y}=\frac{B}{\alpha }\left[ 
\frac{\pi }{2}\left( y_{SR}+y_{SL}\right) -\pi \left( y_{SR}.y_{SL}\right)
^{1/2}\right] .  \label{morse2}
\end{equation}

Combining this result with the preceding expressions for the turning points,
we readily obtain%
\begin{equation}
I^{SWKB}=\frac{B}{\alpha }\left[ \frac{\pi }{2}\left( y_{PR}+y_{PL}\right)
-\pi \left( y_{PR}.y_{PL}\right) ^{1/2}-\frac{\pi }{2}\right] .
\end{equation}

Using%
\begin{equation}
\frac{\pi }{2}\left( y_{PR0}+y_{PL0}\right) -\pi \left(
y_{PR0}.y_{PL0}\right) ^{1/2}=\left( \alpha /B\right) \frac{\pi }{2},
\end{equation}%
and the integral formula 
\begin{equation}
I^{PQ}=\left( B/\alpha \right) \dint_{y_{SL}}^{y_{SR}}\sqrt{\left(
y_{PR}-y\right) \left( y-y_{PL}\right) }\frac{dy}{y},
\end{equation}%
we then obtain%
\begin{equation}
I^{SWKB}=I^{PQ}\left( E_{n}\right) -I^{PQ}\left( E_{0}\right) .
\end{equation}

\subsection{Kepler-Coulomb potential}

The Kepler-Coulomb potential with zero-energy ground state is \cite{6}, \cite%
{11}%
\begin{equation}
V(x)=-\frac{\gamma }{x}+\frac{l(l+1)}{x^{2}}+\frac{\gamma ^{2}}{4(l+1)^{2}}.
\end{equation}

Introducing the variable $y=1/x$, it becomes $V(y)=-\gamma
y+l(l+1)y^{2}+\gamma ^{2}/4(l+1)^{2}$ and the corresponding superpotential
is given by \cite{9} $W\left( y\right) =-a/x+b,$ \ where $a=l+1$ $>0$ and $%
b=\gamma /4(l+1)>0$ .

The SWKB turning points in terms of the $x$ variable satisfy

\begin{equation}
\left\{ 
\begin{array}{c}
x_{SR}+x_{SL}=2ab/\left( b^{2}-E_{n}^{(-)}\right) \\ 
x_{SR}.x_{SL}=a^{2}/\left( b^{2}-E_{n}^{(-)}\right)%
\end{array}%
\right.
\end{equation}%
and those for the proper quantization 
\begin{equation}
\left\{ 
\begin{array}{c}
x_{PR}+x_{PL}=2ab/\left( b^{2}-E_{n}\right) \\ 
x_{PR}.x_{PL}=a(a-1)/\left( b^{2}-E_{n}\right) .%
\end{array}%
\right.
\end{equation}

The integral for the SWKB formula in term of the $x$ variable is written as%
\begin{equation}
I^{SWKB}=\sqrt{b^{2}-E_{n}^{(-)}}\dint_{y_{SL}}^{y_{SR}}\sqrt{\left(
x_{SR}-x\right) \left( x-x_{SL}\right) }\frac{dx}{x}.
\end{equation}

Using the above expression of the different turning points, we show easily
that 
\begin{equation}
\frac{\pi }{2}\left( x_{SR}+x_{PL}\right) -\pi \left( x_{SR}.x_{SL}\right)
^{1/2}=\frac{\pi }{2}\left( x_{PR}+x_{PL}\right) -\pi \left(
x_{PR}.x_{PL}\right) ^{1/2}-\pi a/b+\pi a(a-1)/b,
\end{equation}%
and

\begin{equation}
\frac{\pi }{2}\left( x_{PR0}+x_{PL0}\right) -\pi \left(
x_{PR0}.x_{PL0}\right) ^{1/2}=\pi a/b-\pi a(a-1)/b.
\end{equation}

Since we have also%
\begin{equation}
I^{PQ}=\sqrt{b^{2}-E_{n}}\dint_{y_{SL}}^{y_{SR}}\sqrt{\left( x_{SR}-x\right)
\left( x-x_{SL}\right) }\frac{dx}{x},
\end{equation}%
we directly deduce%
\begin{equation}
I^{SWKB}=I^{PQ}\left( E_{n}\right) -I^{PQ}\left( E_{0}\right) .
\end{equation}

\subsection{Isotonic potential}

The isotonic potentiel with zero-energy ground state is \cite{6}, \cite{11}%
\begin{equation}
V(x)=\frac{\omega ^{2}}{4}x^{2}+\frac{l(l+1)}{x^{2}}-\omega \left( l+\frac{3%
}{2}\right) ,
\end{equation}%
has an associated superpotential of the form \cite{9} $W\left( x\right)
=ax-b/x,$ where $a=\omega /2>0$ and $b=l+1>0$.

The SWKB turning points in terms of $y=x^{2}$ are given by 
\begin{equation}
\left\{ 
\begin{array}{c}
y_{SR}+y_{SL}=\left( 2ab+E_{n}^{(-)}\right) /a^{2} \\ 
y_{SR}.y_{SL}=b^{2}/a^{2},%
\end{array}%
\right.
\end{equation}%
and those for the proper quantization are given by : 
\begin{equation}
\left\{ 
\begin{array}{c}
y_{PR}+y_{PL}=\left( E_{n}+\omega \left( l+3/2\right) \right) /a^{2} \\ 
y_{PR}.y_{PL}=b(b-1)/a^{2}.%
\end{array}%
\right.
\end{equation}

In terms of the $y$ variable, the SWKB integral takes the form 
\begin{equation*}
I^{SWKB}=\frac{a}{2}\dint_{y_{SL}}^{y_{SR}}\sqrt{\left( y_{SR}-y\right)
\left( y-y_{SL}\right) }\frac{dy}{y}=\pi \frac{E_{n}^{(-)}}{2\omega }.
\end{equation*}

Similarly, we have also

\begin{equation*}
\ I^{PQ}=\frac{a}{2}\dint_{y_{SL}}^{y_{SR}}\sqrt{\left( y_{PR}-y\right)
\left( y-y_{PL}\right) }\frac{dy}{y}=\pi \left( E_{n}^{(-)}/2\omega +\left(
l+\frac{3}{2}\right) /2\right) ,
\end{equation*}%
from which we can directly conclude that

\begin{equation}
I^{SWKB}=I^{PQ}\left( E_{n}\right) -I^{PQ}\left( E_{n=0}\right) ,
\end{equation}%
for the isotonic potential.

\section{Energy spectra of the PTSIP}

\subsection{First category PTSIP}

For the first category of potentials, the calculation of the energy spectrum
is achieved by solving equation $I^{SWKB}=n\pi $, where $I^{SWKB}$ is given
by $\left( \ref{17}\right) $. Consider the two equations in $\left( \ref{17}%
\right) $ and introduce the notations $\eta _{1}=E_{n}^{-}+a^{2}-b^{2}$ into
the first equation and $\eta _{2}=\sqrt{\left( a+b\right) ^{2}-E_{n}^{-}}$
into the second one. The substitution of these variables into the identity $%
I^{SWKB}=n\pi $ gives correspondingly

\begin{equation}
\left\{ 
\begin{array}{c}
\sqrt{\eta _{1}^{2}+4a^{2}b^{2}}+\eta _{1}=2\left( a+\alpha n\right) ^{2} \\ 
-\sqrt{\eta _{2}^{2}-4a^{2}b^{2}}-\eta _{2}=2\left( \alpha n-a\right) \text{
.}%
\end{array}%
\right.
\end{equation}

Since $\xi =2\left( a+\alpha n\right) ^{2}$ is one of the two solutions of
the quadratic equation $(1/2)\xi ^{2}-\eta _{1}\xi -2a^{2}b^{2}=0$, it is
sufficient to solve the latter with respect to $\eta _{1}$. Following the
same method to find $\eta _{2}$, one obtaines the analytical expression of
energy levels%
\begin{equation}
E_{n}=\mp a^{2}+\alpha a+\lambda _{0}\pm \left( a\pm \alpha n\right) ^{2}-%
\frac{\lambda _{1}^{2}}{4\left( a\pm \alpha n\right) ^{2}}.
\end{equation}

Equation $\left( \ref{11}\right) $ allows us to express the energy as a
function of the parameters $\lambda _{2},\lambda _{1},$ $\lambda _{0}$ and $%
\alpha $. To illustrate this result, we consider the Rosen-Morse I and II
potentials and the Eckart potential.

\subsubsection{\textbf{Rosen-Morse I}}

The Rosen--Morse I potential with zero-energy ground state has the form \cite%
{6}, \cite{11} 
\begin{equation}
V(x)=\frac{A(A-\alpha )}{\sin ^{2}(\alpha x)}+2B\cot \left( \alpha x\right)
-A^{2}+\frac{B^{2}}{A^{2}}\text{, }0\leq \alpha x\leq \pi .  \label{26}
\end{equation}

Using the variable $y=-\cot \left( \alpha x\right) $ ($y^{\prime }(x)=\alpha
\left( 1+y^{2}\right) $), it becomes$\ V(y)=A(A-\alpha
)y^{2}-2By+B^{2}/A^{2}-\alpha A$ . Comparing with $\left( \ref{7}\right) $
gives%
\begin{equation}
a=A,\text{ }b=-\frac{B}{A},  \label{27}
\end{equation}%
from which, we deduce for the superpotential $W\left( x\right) =ay+b=A(-\cot
\left( \alpha x\right) )-$ $B/A$ and for the energy levels%
\begin{equation}
E_{n}=-A^{2}+\frac{B^{2}}{A^{2}}+\left( A+\alpha n\right) ^{2}-\frac{B^{2}}{%
\left( A+\alpha n\right) ^{2}}.  \label{28}
\end{equation}

\subsubsection{\textbf{Rosen-Morse II}}

The Rosen--Morse II potential with zero-energy ground state is given by \cite%
{6}, \cite{11}%
\begin{equation}
V(x)=-\frac{A(A+\alpha )}{\cosh ^{2}(\alpha x)}+2B\tanh \left( \alpha
x\right) +A^{2}+\frac{B^{2}}{A^{2}}\text{, \ }B<A^{2}.  \label{29}
\end{equation}

Using the variable $y=\tanh \left( \alpha x\right) $ ($y^{\prime }(x)=\alpha
\left( 1-y^{2}\right) $), it becomes$\ V(y)=A(A+\alpha
)y^{2}+2By+B^{2}/A^{2}-\alpha A$. Comparing with $\left( \ref{7}\right) $
gives%
\begin{equation}
a=A,b=\frac{B}{A},
\end{equation}%
which leads to a superpotential $W\left( x\right) =A\tanh \left( \alpha
x\right) +$ $B/A$ and the energy spectrum is%
\begin{equation}
E_{n}=A^{2}+\frac{B^{2}}{A^{2}}-\left( A-\alpha n\right) ^{2}-\frac{B^{2}}{%
\left( A-\alpha n\right) ^{2}}.  \label{30}
\end{equation}

\subsubsection{\textbf{Eckart potential }}

The Eckart potential with zero-energy ground state is given by \cite{6}, 
\cite{11}:%
\begin{equation}
V(x)=\frac{A(A-\alpha )}{\sinh ^{2}(\alpha x)}-2B\coth \left( \alpha
x\right) +A^{2}+\frac{B^{2}}{A^{2}}\text{, }B>A^{2}.  \label{31}
\end{equation}

Using the variable $y=\coth \left( -\alpha x\right) $ $\ $and redefining $%
\alpha \rightarrow -\alpha $, it becomes$\ V(y)=A(A+\alpha
)y^{2}+2By+B^{2}/A^{2}-\alpha A$ which is exactly the form obtained in the
previous case. The superpotential is then $W\left( x\right) =A\coth \left(
-\alpha x\right) +$ $B/A$ and the energy levels are simply obtained by
replacing $\alpha $ by $-\alpha $ in $\left( \ref{30}\right) $%
\begin{equation}
E_{n}=A^{2}+\frac{B^{2}}{A^{2}}-\left( A+\alpha n\right) ^{2}-\frac{B^{2}}{%
\left( A+\alpha n\right) ^{2}}.  \label{32}
\end{equation}

\subsection{Second category PTSIP}

For the second category of potentials, the resolution of $I^{SWKB}=n\pi $
leads to the following expression for energy levels \cite{kasri2015}

\begin{equation}
E_{n}=\lambda _{0}\mp \left( \mu _{2}+\lambda _{2}\right) \pm \left( 2\alpha
n\pm a+b\right) ^{2}.  \label{33}
\end{equation}

\subsubsection{\textbf{P\"{o}schl-Teller}}

The P\"{o}schl--Teller potential with zero-energy ground state is given by 
\cite{6}, \cite{11}

\begin{equation}
V(x)=A^{2}+\frac{A^{2}+B^{2}+\alpha A}{\sinh ^{2}(\alpha x)}-B(2A+\alpha )%
\frac{\coth \left( \alpha x\right) }{\sinh (\alpha x)},\quad B>A,\quad x>0.
\label{34}
\end{equation}

The choice of variable $y=\tanh \left( \alpha x/2\right) $ transforms it
into $V(y)=a(a+\alpha /2)y^{2}-b(b-\alpha /2)/y^{2}+\lambda _{0},$ where 
\begin{equation}
a=\frac{A+B}{2},\quad b=\frac{B-A}{2}
\end{equation}%
and $\lambda _{0}=-\alpha \left( a-b\right) /2-2ab$. In this case $y^{\prime
}(x)=\alpha \left( 1-y^{2}\right) /2$, and one obtains as superpotential $%
W\left( x\right) =ay-b/y=A\coth \left( \alpha x\right) -B/\sinh (\alpha x)$
and as energy spectrum (see $\left( \ref{33}\right) )$ 
\begin{equation}
E_{n}=A^{2}-\left( \alpha n-A\right) ^{2}.  \label{35}
\end{equation}

\subsubsection{\textbf{P\"{o}schl-Teller I}}

The P\"{o}schl--Teller I potential with zero-energy ground state has the
form \cite{6}, \cite{11}

\begin{equation}
V(x)=-\left( A+B\right) ^{2}+\frac{A(A-\alpha )}{\cos ^{2}(\alpha x)}+\frac{%
B(B-\alpha )}{\sin ^{2}(\alpha x)},\text{ \ }0<x<\frac{\pi }{2\alpha }.
\end{equation}

Introducing the variable $y=\tan \alpha x$ ($y^{\prime }(x)=\alpha \left(
1+y^{2}\right) $), this potential takes the form $V(y)=a(a-\alpha
)y^{2}-b(b-\alpha )/y^{2}+\lambda _{0},$ where 
\begin{equation}
a=A,b=B,
\end{equation}%
and $\lambda _{0}=-\alpha \left( a+b\right) -2ab$. One obtains $W\left(
x\right) =ay-b/y=A\tan \left( \alpha x\right) -B\cot (\alpha x)$ and the
energy spectrum is (see $\left( \ref{33}\right) $) 
\begin{equation}
E_{n}=-\left( A+B\right) ^{2}+\left( 2\alpha n+A+B\right) ^{2}.
\end{equation}

\subsubsection{\textbf{Poschl-Teller II }}

The P\"{o}schl--Teller II potential with zero-energy ground state has the
following form \cite{6}, \cite{11}%
\begin{equation}
V(x)=\left( A-B\right) ^{2}-\frac{A(A+\alpha )}{\cosh ^{2}(\alpha x)}+\frac{%
B(B-\alpha )}{\sinh ^{2}(\alpha x)},\text{ }B<A,x>0.
\end{equation}

By using the variable $y=\tanh \alpha x$, we obtain $V(y)=a(a+\alpha
)y^{2}+b(b-\alpha )/y^{2}+\lambda _{0},$ where 
\begin{equation}
a=A,b=B,
\end{equation}%
and $\lambda _{0}=-\alpha \left( a-b\right) -2ab$. The superpotential
becomes $W\left( x\right) =A\tanh \left( \alpha x\right) -B\coth (\alpha x)$
and the energy levels are given by 
\begin{equation}
E_{n}=\left( B-A\right) ^{2}-\left( 2\alpha n+B-A\right) ^{2}.
\end{equation}

\subsubsection{\textbf{Scarf I}}

The Scarf I potential with zero-energy ground state is given by \cite{6}, 
\cite{9}%
\begin{equation}
V(x)=-A^{2}+\frac{A^{2}+B^{2}-\alpha A}{\sin ^{2}(\alpha x)}-B(2A-\alpha )%
\frac{\cot \left( \alpha x\right) }{\sin (\alpha x)},\text{ }B<A,0<x<\frac{%
\pi }{2\alpha },
\end{equation}

With the introduction of the variable $y=\tan \left( \alpha x/2\right) $, it
takes the form $V(y)=a(a-\frac{\alpha }{2})y^{2}+b(b-\alpha
/2)/y^{2}+\lambda _{0}$,\ where 
\begin{equation}
a=\frac{A+B}{2},b=\frac{A-B}{2}
\end{equation}%
and $\lambda _{0}=-\alpha \left( a+b\right) /2-2ab$. In this case $y^{\prime
}(x)=\alpha \left( 1+y^{2}\right) /2$ and one obtains the superpotential $%
W\left( x\right) =-A/\tan \left( \alpha x\right) +B/\sin (\alpha x)$. The
energy levels are deduced from $\left( \ref{33}\right) $ as%
\begin{equation}
E_{n}=-A^{2}+\left( A+\alpha n\right) ^{2}.
\end{equation}

\section{Conclusion}

In this work, we have examined the link between the Ma-Xu or proper
quantization rule and the SWKB condition. By using the classification into
two categories of the PTSIP, we have shown the equivalence of these two
quantization rules. The study covers the whole set of PTSIP, including the
exceptional cases. As an application of the general classification
introduced in \cite{9}, a global formula for the energy levels of every
PTSIP has been extracted from the SWKB, recovering in each specific case the
known results \cite{7,8}.

A challenging problem for the future would be to establish exact
quantization formulas for the solvable potentials obtained as rational
extensions of the PTSIP considered above \cite%
{52,grandati,quesne,gomez,ramos,sasaki}. It seems indeed that the SWKB
quantization formula is not exact for these potentials even for those which
belong to the class of TSIP \cite{81}. This has to be correlated to the $%
\hbar $ dependence of the superpotentials associated to these extensions in
contrast with the primary TSIP for which the superpotential does not present
such dependence \cite{bougie}.

\section{Appendix 1 - Evaluation of $I^{PQ}\left( E_{0}\right) $}

\subsection{\textbf{First category }}

Taking $n=0$ in the first terms of the two equations of $\left( \ref{20}%
\right) $, one finds

\begin{equation}
I^{PQ}\left( E_{0}\right) =\left\{ 
\begin{array}{c}
\frac{1}{\alpha }\left[ \frac{\pi }{\sqrt{2}}\left( \sqrt{\left(
E_{0}-\lambda _{0}+\lambda _{2}\right) ^{2}+\lambda _{1}^{2}}+E_{0}-\lambda
_{0}+\lambda _{2}\right) ^{\frac{1}{2}}-\sqrt{\lambda _{2}}\pi \right] \text{%
, \ if }y^{\prime }/\alpha =1+y^{2}, \\ 
\\ 
\frac{\pi }{2\alpha }\left[ 2\sqrt{\lambda _{2}}-\sqrt{\lambda _{2}+\lambda
_{1}+\lambda _{0}-E_{0}}-\sqrt{\lambda _{2}-\lambda _{1}+\lambda _{0}-E_{0}}%
\right] \text{, \ if }y^{\prime }/\alpha =1-y^{2}.%
\end{array}%
\right.  \label{app 1}
\end{equation}

Using $\left( \ref{9}\right) $ to $\left( \ref{11}\right) $, it is
straightforward show that 
\begin{equation}
I^{PQ}\left( E_{0}\right) =\left\{ 
\begin{array}{c}
\left( \sqrt{\left( E_{0}-\lambda _{0}+\lambda _{2}\right) ^{2}+\lambda
_{1}^{2}}+E_{0}-\lambda _{0}+\lambda _{2}\right) ^{\frac{1}{2}}=\sqrt{2}a%
\text{, \ if }y^{\prime }=\alpha \left( 1+y^{2}\right) , \\ 
\sqrt{\lambda _{2}\pm \lambda _{1}+\lambda _{0}-E_{0}}=\left\vert a\pm
b\right\vert \text{, \ \ \ \ \ \ \ \ \ \ \ \ \ \ \ \ \ \ \ \ \ \ \ \ if }%
y^{\prime }=\alpha \left( 1-y^{2}\right) ,%
\end{array}%
\right.
\end{equation}%
and inserting these identities into $\left( \ref{app 1}\right) $\ we obtain%
\begin{equation}
I^{PQ}\left( E_{0}\right) =\left\{ 
\begin{array}{c}
\frac{\pi }{\alpha }\left( a-\sqrt{\lambda _{2}}\right) ,\text{ \ \ \ \ \ \
\ \ \ \ \ if }y^{\prime }/\alpha =1+y^{2}, \\ 
\\ 
-\frac{\pi }{\alpha }\left( a-\sqrt{\lambda _{2}}\right) \text{, }\left(
b<a\right) \text{\ }\ \ \text{ \ \ if }y^{\prime }/\alpha =1-y^{2},\text{\ }%
\end{array}%
\right.  \label{app 4}
\end{equation}%
which is the desired result.

\subsection{\textbf{Second category}}

As we have seen, the first terms of the right-hand side of equation $\left( %
\ref{200}\right) $ correspond to $I^{PQ}\left( E_{n}\right) $. For $n=0$,
the expression of $I^{PQ}\left( E_{0}\right) $ is written under the
following form

\begin{equation}
I^{PQ}\left( E_{0}\right) =\frac{\sqrt{\lambda _{2}}}{2\alpha }\pi \left[
\left( \mp 1-\sqrt{\frac{\mu _{2}}{\lambda _{2}}}\pm \sqrt{1+\frac{\mu _{2}}{%
\lambda _{2}}\pm \frac{E_{0}-\lambda _{0}}{\lambda _{2}}}\right) \right] .
\end{equation}

By means of $\left( \ref{15}\right) $ and taking $n=0$, one gets%
\begin{equation}
\left( b\pm a\right) ^{2}=\pm E_{0}\mp \lambda _{0}+\lambda _{2}+\mu _{2}.
\end{equation}

Substituting the last relation in $\left( \ref{18}\right) $ leads to%
\begin{equation}
I^{PQ}\left( E_{0}\right) =\frac{\pi }{2\alpha }\left( b\pm a\mp \sqrt{%
\lambda _{2}}-\sqrt{\mu _{2}}\right) ,
\end{equation}%
where $b<a$ in the second case, i.e. when $y^{\prime }\left( x\right)
=\alpha \left( 1-y^{2}\right) $.

\section{Appendix 2}

For the first category of potentials, the following integral formulas \cite%
{10} have been used 
\begin{eqnarray}
\dint_{\alpha }^{\beta }\sqrt{\left( \beta -y\right) \left( y-\alpha \right) 
}\frac{dy}{1+y^{2}} &=&\frac{\pi }{\sqrt{2}}\left[ \sqrt{1+\alpha ^{2}}\sqrt{%
1+\beta ^{2}}-\alpha \beta +1\right] ^{1/2}-\pi ,\alpha <\beta ,  \label{13a}
\\
\dint_{\alpha }^{\beta }\sqrt{\left( \beta -y\right) \left( y-\alpha \right) 
}\frac{dy}{1-y^{2}} &=&\frac{\pi }{2}\left[ 2-\sqrt{\left( 1-\alpha \right)
\left( 1-\beta \right) }-\sqrt{\left( 1+\alpha \right) \left( 1+\beta
\right) }\right] ,\text{ }  \label{13b}
\end{eqnarray}%
where $-1<\alpha <\beta <1$ for the second formula. For the second category
of potentials, we have the following formula \cite{kasri2015}

\begin{equation}
\int_{\alpha }^{\beta }\sqrt{\left( \beta -z\right) \left( z-\alpha \right) }%
\frac{dz}{z\left( 1\pm z\right) }=\pi \left[ \mp 1-\sqrt{\alpha \beta }\pm 
\sqrt{\left( 1\pm \alpha \right) \left( 1\pm \beta \right) }\right] .
\label{15c}
\end{equation}

The signs $\pm $ in the last integral represents two different cases and the
formula is valid under the following condition: the parameters $\alpha $ and 
$\beta $\ are positive in both cases and are lower than unity in the second.

\end{document}